\documentclass[cameraready]{Interspeech}

\title{IRAF: Interference-Resilient Adaptive Fusion for Noise-Robust End-to-End Full-Duplex Spoken Dialogue Systems}

\author[affiliation={1, 2}, , ]{Tao}{Zhong}
\author[affiliation={2}, , , correspondingauthor]{Jiajun}{Deng}
\author[affiliation={3}, ,]{Nikita}{Kuzmin}
\author[affiliation={2}]{Yinke}{Zhu}
\author[affiliation={2}]{Tianxiang}{Cao}
\author[affiliation={2}]{Tristan}{Tsoi}
\author[affiliation={2}]{Zhili}{Tan}
\author[affiliation={2}]{Simon}{Lui}
\author[affiliation={1}, , correspondingauthor]{Xunying}{Liu}

\address{
    $^1$ The Chinese University of Hong Kong, China \\
    $^2$ AudioLab Hong Kong, Huawei Leibniz Research Center, China \\
    $^3$ Nanyang Technological University, Singapore
}

\email{tzhong@se.cuhk.edu.hk, jjdeng321@gmail.com, 
xyliu@se.cuhk.edu.hk}


\usepackage{comment}

\usepackage{xurl}
\usepackage{multirow}
\usepackage{svg}
\usepackage{subfigure}
\usepackage{subcaption}
\usepackage{quotes}
\usepackage{tabularx}
\usepackage[table]{xcolor} 
\usepackage{colortbl}
\usepackage{quoting}
\usepackage{cite}

\begin{document}

\maketitle

\begin{abstract}
Full-duplex spoken dialogue models allow voice agents to listen and speak concurrently, enabling natural interaction with real-time overlap. However, end-to-end dual-channel models that jointly encode user and agent streams may degrade in realistic acoustic environments: interfering speakers leaking into the user microphone can be encoded as part of the user query, corrupting the LLM’s conditioning and causing unstable turn-taking and reduced response quality. We propose Interference-Resilient Adaptive Fusion (IRAF), a lightweight, streaming-compatible module that modulates the contribution of user audio to the LLM frame by frame. IRAF predicts a scalar reliability gate from target-speaker and user audio embeddings and rescales user representations before fusion with agent embeddings. Experiments on MS-MARCO and InstructS2S-200K show consistent gains in response quality and full-duplex interaction under interfering-speaker conditions.

\end{abstract}

\section{Introduction}
Recent advances in voice agents have shifted attention toward full-duplex spoken dialogue models that can listen and speak concurrently and manage conversational overlap in real time. Compared to conventional turn-based voice agents~\cite{turn_base,zhang-etal-2023-speechgpt,kim2024paralinguisticsaware,zeng2024glm4,Audio_gpt,du2024lauragptlistenattendunderstand,10613503,fang2025llamaomni}, full-duplex capability supports continuous, natural interaction that more closely resembles human conversation.

Existing approaches to full-duplex spoken dialogue models can be broadly categorized into two lines of work. \textbf{a) Modular, controller-style systems}~\cite{wang2024full,liao2025flexduopluggableenablingfullduplex,chen2025fireredchatpluggablefullduplexvoice,chen2025minmomultimodallargelanguage,xiong2024freeze,xie2024miniomni2opensourcegpt4ovision,zhang2025llm,speak_listen,li2025easyturnintegratingacoustic}, for example, Neural FSM~\cite{wang2024full}, FlexDuo~\cite{liao2025flexduopluggableenablingfullduplex}, FireRedChat~\cite{chen2025fireredchatpluggablefullduplexvoice}, and semantic VAD-based dialogue managers~\cite{zhang2025llm}, introduce explicit control signals to coordinate when the agent should listen, speak, or stop. In contrast, \textbf{b) End-to-end (E2E) native dual-channel} speech language models~\cite{moshi,yu2025salmonn,wang2025ntpp,zufle2026f,hu25f_interspeech,wu2025chronologicalthinkingfullduplexspoken,ohashi25_interspeech,chen2025reinforcement,roy2026personaplexvoicerolecontrol}, for example, Moshi~\cite{moshi}, SALMONN-omni~\cite{yu2025salmonn}, NTPP~\cite{wang2025ntpp}, and F-Actor~\cite{zufle2026f}, learn to model user and agent streams jointly, often with additional mechanisms for synchronization such as external stop commands or time-chunking strategies that align audio and text over time. Recent work further simplifies duplex modeling by reducing reliance on speech pretraining (e.g., SALM-Duplex~\cite{hu25f_interspeech}) or by employing transformer decoders~\cite{wu2025chronologicalthinkingfullduplexspoken} to directly generate response audio, highlighting a trend toward more unified, LLM-centric duplex architectures. 

\begin{figure}[!t]
  \centering
  \includegraphics[width=\linewidth]{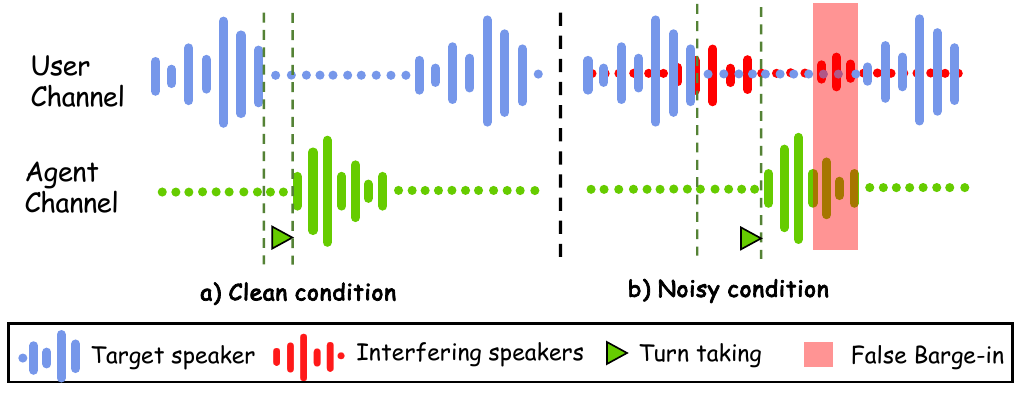}
  \vspace{-0.6cm}
  \caption{Full-duplex dialogue in (a) clean and (b) noisy conditions. Interference leaking into the user channel can corrupt conditioning, causing unstable turn-taking and false barge-in.}
  \label{fig:examples}
  \vspace{-0.5cm}
\end{figure}

\begin{figure*}[!t]
  \centering
  \includegraphics[width=\linewidth]{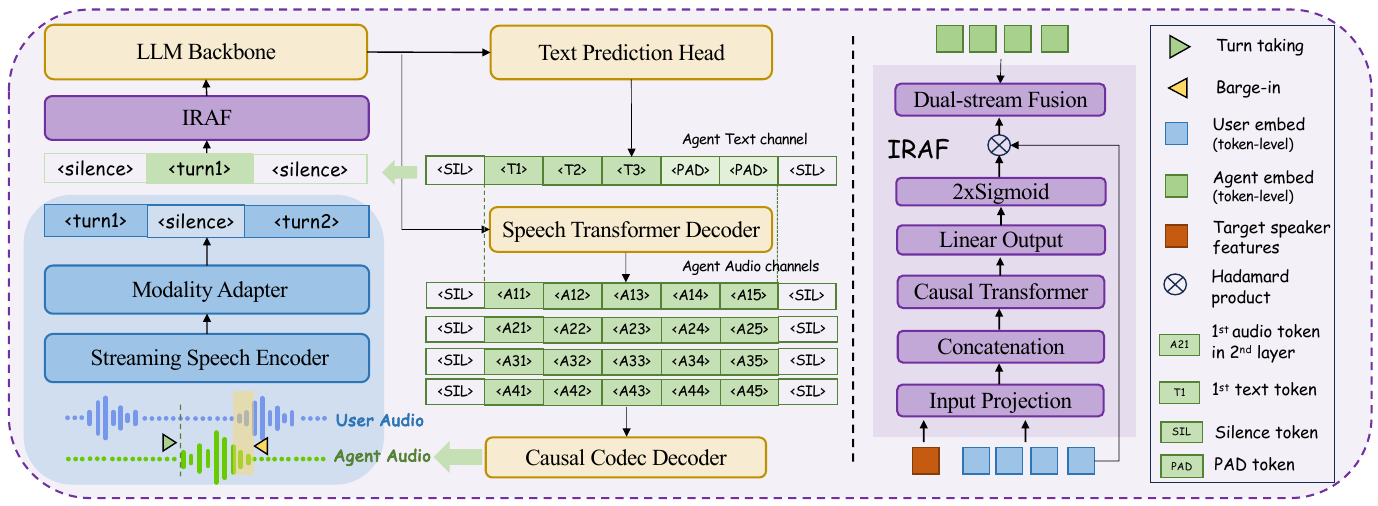}
  \vspace{-0.5cm}
  \caption{Overview of the proposed E2E full-duplex model with the Interference-Resilient Adaptive Fusion (IRAF) module. A streaming speech encoder produces frame-level user embeddings, which IRAF adaptively gates before fusion with agent text embeddings and processing by the LLM to generate text tokens; a speech decoder, conditioned on the LLM hidden states, generates audio tokens.}
  \label{fig:baseline_duplex_system}
  \vspace{-0.5cm}
\end{figure*}

While native dual-channel multi-stream models benefit from end-to-end optimization and often perform well in clean conditions, their robustness can degrade substantially in real-world acoustic environments. For instance, when interfering speakers leak into the user microphone, their speech can be encoded into the user stream and inadvertently treated as part of the user query, thereby contaminating the LLM’s conditioning context. This misconditioning can disrupt both interaction control (e.g., unstable turn-taking and mistimed responses) and reasoning performance, shown in Figure~\ref{fig:examples}. 

Efforts on developing noise-robust end-to-end full-duplex modeling are confronted with several challenges. First, the model operates under strict causality: it must perceive user audio and generate responses incrementally, without lookahead or access to a complete utterance. Second, interference is highly non-stationary, and overlap is largely unconstrained \cite{miles2023behavioral}. For example, background speakers may appear abruptly and coincide with the target user, requiring the model to handle rapidly changing acoustic conditions. Third, duplex interaction imposes tight latency constraints; the agent must respond immediately when the user yields the floor, leaving limited budget for computationally expensive separation or post-hoc filtering. 

To this end, we develop an end-to-end full-duplex spoken dialogue model that is robust to realistic acoustic conditions, especially the presence of interfering speakers. Specifically, we present a lightweight, streaming-compatible Interference-Resilient Adaptive Fusion (IRAF) module, which dynamically modulates the contribution of user audio to the LLM input on a frame-by-frame basis. IRAF combines a target-speaker embedding with the user audio embedding to predict a scalar reliability gate at each time step and uses this gate to rescale the user audio representation before fusing it with the agent-side text embeddings. Experiments on the open-source single-turn MS MARCO conversational dataset~\cite{bajaj2018msmarcohumangenerated} and the multi-turn InstructS2S-200K conversational dataset~\cite{fang2025llamaomni2} show that the proposed end-to-end duplex model consistently outperforms strong baselines across a range of metrics, including response quality (ASR-based lexical and semantic measures) and full-duplex interaction (turn-taking and barge-in performance). The main contributions of this paper are as follows:

1) To the best of our knowledge, this is the first attempt to address noise- and interference-induced conditioning corruption in end-to-end full-duplex spoken dialogue systems. In contrast, prior E2E full-duplex models have largely focused on duplex behavior under relatively clean conditions~\cite{wang2025ntpp,zufle2026f,hu25f_interspeech,wu2025chronologicalthinkingfullduplexspoken,ohashi25_interspeech} or on mitigating agent echo~\cite{yu2025salmonn} and have not systematically tackled complex acoustic environments.

2) This paper proposes a lightweight, streaming-compatible Interference-Resilient Adaptive Fusion (IRAF) module that preserves the end-to-end formulation without introducing additional response latency. IRAF estimates a frame-level reliability gate from target-speaker embeddings as well as the user channel and uses it to modulate the user audio representation before fusing it with the agent channel. The effectiveness of the proposed method is demonstrated through extensive evaluations.

\vspace{-0.1cm}
\section{Interference-Resilient Duplex Modeling}
\vspace{-0.2cm}
This section presents an end‑to‑end full‑duplex speech dialogue model equipped with an Interference-Resilient Adaptive Fusion (IRAF) mechanism that enables the model to process speech input and generate responses in parallel, thereby enabling natural, overlapping interaction under noisy conditions. 


\vspace{-0.1cm}
\subsection{Multi-stream Duplex Modeling}
\vspace{-0.15cm}
The network architecture is depicted in Figure~\ref{fig:baseline_duplex_system}. It consists of two synchronized input streams: a user speech stream and an agent text stream. More specifically, the user’s audio signal is processed by a streaming speech encoder operating at 12.5 Hz, along with a modality adapter that converts the acoustic waveform into a sequence of continuous embeddings \( X \in \mathbb{R}^{T\times D} \), where \( T \) and \(D\) denote the total number of frames and the dimension of the audio embedding, respectively. The resulting user audio embeddings are then combined with the agent’s text embeddings \( Y^{txt} \in \mathbb{R}^{T\times D} \) before being passed to the LLM backbone. Following the work \cite{wu2025chronologicalthinkingfullduplexspoken}, to reduce the computational complexity of the LLM backbone while preserving contextual coherence, a separate autoregressive speech transformer decoder is built on top of the LLM backbone to predict the agent's speech tokens \(Y^{a} \in \mathbb{R}^{T}\) conditioned on the last hidden states of LLM \( h \). All the model parts (except the Causal Codec Decoder) are jointly fine‑tuned under a multi‑channel next‑token prediction objective, enabling E2E optimization across both speech and text modalities, given as  
\vspace{-0.3cm}
\begin{equation}
\begin{aligned}
   \mathcal{L}(Y^{txt},Y^{a}|X,\theta, \phi) 
   &= - \sum_{t=1}^{T} \{ \lambda_1 \log p_{\theta}(Y_t^{txt} \mid Y_{<t}^{txt}, X) \\
   & + \lambda_2 \log p_{\phi}(Y_{t}^{a} \mid Y_{<t}^{a}, h_{<t}) \}, \label{eq1:loss}
\end{aligned}
\end{equation}
where $\theta$ and $\phi$ denote the parameters of LLM backbone and the speech transformer decoder, respectively. $\lambda_1$ and $\lambda_2$ are the agent text and audio cross-entropy loss weights respectively, which are set to 1.0 and 5.0 in this paper. The agent speech tokens are extracted from the Nanocodec with finite scalar quantization of 12.5 Hz. 

\vspace{-0.2cm}
\subsection{Interference-Resilient Adaptive Fusion}
\vspace{-0.15cm}
In the duplex multi-stream modeling, the LLM is conditioned on a single fused representation obtained by directly summing the user audio embedding and the agent text embedding. This design is simple, but it treats the audio stream as uniformly trustworthy over time. In real-world acoustic scenes, that assumption is frequently violated: when background speakers or other interference are present, portions of the user audio embedding may primarily reflect non-target speech. Because the fusion is unconditional, these corrupted frames are injected into the LLM in the same way as clean target-user speech, contaminating the conditioning signal and often leading to unstable or incorrect responses.

To make the conditioning robust to such interference, we propose a simple yet effective Interference-Resilient Adaptive Fusion (IRAF) method. Instead of always mixing the user audio with a fixed weight, IRAF predicts a frame-level reliability gate indicating whether the current acoustic evidence is consistent with the target speaker and then uses this estimate to scale the audio contribution before being passed to the LLM. As a result, interference-dominated frames contribute minimally to the LLM input, while frames containing target-user speech are preserved, yielding more stable behavior in noisy, multi-speaker settings. Formally, at each time step \(t\), we concatenate the target-speaker embedding \(s \in \mathbb{R}^{n}\) with the user audio embedding \(X_t \in \mathbb{R}^{D}\), and feed the resulting representation into a transformer-based fusion module \(f(\cdot \mid \psi)\). The module comprises (i) an input projection block that jointly maps speaker and audio features into a shared space, (ii) a causal Transformer layer that aggregates streaming context, and (iii) a linear output layer that produces the reliability estimate, given by
\vspace{-0.2cm}
\begin{equation}
g_t = 2\times\text{Sigmoid}(f(s, X_{\leq t}|\psi))\in [0,2],
\end{equation}
where the scalar gate $g_t$ is used to rescale the user audio embedding and sum with the agent text embedding before feeding the result to the text LLM, given as $g_t*X_t + Y_t^{txt}$.

IRAF is trained jointly with the full speech dialogue model in an end-to-end manner. From clean training utterances, we derive frame-level targets by labeling frames as 1 when the target speaker is active and 0 otherwise, and add an auxiliary binary classification loss (weight 0.1) to the main objective in Eqn.~(\ref{eq1:loss}).

\vspace{-0.3cm}

\section{Full-Duplex Dataset Generation} \label{sec:experiments}
\vspace{-0.1cm}
Two publicly available datasets are used to evaluate the proposed method: (a) Single-turn \textbf{MS MARCO}\footnote{\url{https://microsoft.github.io/msmarco/}}~\cite{bajaj2018msmarcohumangenerated} is a large-scale, single-turn text question answering (QA) benchmark consisting of real anonymized Bing queries paired with human-written answers. Following prior practice for spoken QA, we synthesize speech for these QA pairs using CosyVoice2~\cite{du2024cosyvoice}. (b) Multi-turn \textbf{InstructS2S-200K}\footnote{\url{https://huggingface.co/datasets/ICTNLP/InstructS2S-200K}}~\cite{fang2025llamaomni2} is a multi-turn speech-to-speech dialogue dataset comprising approximately 200,000 conversation sessions. It covers a broad range of interaction types, including common-sense or general world knowledge.

\noindent\textbf{Multi-stream dataset simulation}: To synthesize full-duplex training examples, we convert each conversation session into two time-synchronized streams: a user channel and an agent channel. When one party is speaking, we fill the other channel with silence, yielding non-overlapping duplex signals. In addition, we insert a fixed 0.64s inter-turn pause between the end of the user's utterance and the start of the agent's response.

\noindent\textbf{User interruption dataset simulation}: To simulate user interruptions and the agent’s ability to stop speaking, we augment the multi-turn InstructS2S-200K dataset to create barge-in events. Specifically, we shorten the inter-turn gap in the user stream so that the next user utterance temporally overlaps with the agent’s ongoing speech in the current turn, producing overlap in the duplex training examples. Once the interruption begins, we further simulate an explicit stop behavior by truncating the agent speech and replacing the remaining portion with silence after a fixed latency of 0.64s. During training, we stochastically introduce barge-in events on-the-fly in the data preparation pipeline, which exposes the model to interrupted interactions and improves robustness. Specifically, each turn is modified to include an interruption with probability 0.5.

\noindent\textbf{Real-world full-duplex dataset simulation}:
To approximate real-world operating conditions for full-duplex dialogue, we construct noise-augmented data under two acoustic settings: (a) multi-speaker interference only and (b) multi-speaker interference combined with background noise. We use the MUSAN corpus~\cite{musan2015}, which contains music, speech, and noise recordings. Its speech portion comprises librivox and us-gov subsets. In our simulation, MUSAN speech serves as interfering speakers, and MUSAN noise serves as background noise. To prevent contamination between splits, we divide the MUSAN speech and noise data into three non-overlapping partitions and assign them exclusively to the training, validation, and test sets, respectively. For MS MARCO, we sample the signal-to-noise ratio (SNR) uniformly from 0 to 10dB; for InstructS2S-200K, we use a broader SNR range of 0 to 20dB.

\begin{table}[!htbp]
\centering
\caption{Performance on MS~MARCO under MUSAN speech interference. 
`LIBRI' and `US-GOV' use the corresponding MUSAN speech subsets for test set augmentation, while `ALL' uses all MUSAN speech categories. 
`RL' and `RSR' denote response latency and response success rate, respectively.}
\label{tab:msmarco_people}
\vspace{-0.3cm}

\resizebox{0.95\linewidth}{!}{
\begin{tabular}{llcccc}
\toprule
\multirow{2}{*}{\textbf{Method}} & 
\multirow{2}{*}{\textbf{Noisy Source}} &
\multicolumn{4}{c}{\textbf{Response Performance}} \\ 
\cmidrule(lr){3-6}
& & \textbf{BLEU($\uparrow$)} & \textbf{sBERT($\uparrow$)} & \textbf{RL(s)($\downarrow$)} & \textbf{RSR($\uparrow$)} \\ 
\midrule

\multicolumn{6}{l}{\textit{Interfering speakers only}} \\ 
\midrule
CleanBase & ALL & 0.66 & 0.11 & 1.46 & 6.2\% \\ 
\midrule
\multirow{3}{*}{NoisyAug} 
 & LIBRI & 12.69 & 0.503 & 0.98 & 91.0\% \\
 & US-GOV & 13.30 & 0.512 & 0.96 & 94.1\% \\
 & \cellcolor{cyan!10}ALL 
 & \cellcolor{cyan!10}12.74 
 & \cellcolor{cyan!10}0.506 
 & \cellcolor{cyan!10}0.97 
 & \cellcolor{cyan!10}93.1\% \\ 
\midrule
\multirow{3}{*}{\shortstack{IRAF\\(\textit{Proposed})}} 
 & LIBRI & 13.81 & 0.516 & 0.97 & 95.4\% \\
 & US-GOV & 14.38 & 0.536 & 0.94 & 98.2\% \\
 & \cellcolor{cyan!10}ALL 
 & \cellcolor{cyan!10}\textbf{14.20} 
 & \cellcolor{cyan!10}\textbf{0.523} 
 & \cellcolor{cyan!10}\textbf{0.96} 
 & \cellcolor{cyan!10}\textbf{95.7\%} \\ 
\midrule

\multicolumn{6}{l}{\textit{Both interfering speakers and background noise}} \\ 
\midrule
CleanBase & ALL & 0.00 & 0.03 & 1.49 & 2.8\% \\ 
\midrule
\multirow{3}{*}{NoisyAug} 
 & LIBRI & 11.12 & 0.445 & 0.98 & 87.1\% \\
 & US-GOV & 11.53 & 0.465 & 0.96 & 91.3\% \\
 & \cellcolor{cyan!10}ALL 
 & \cellcolor{cyan!10}11.33 
 & \cellcolor{cyan!10}0.454 
 & \cellcolor{cyan!10}0.98 
 & \cellcolor{cyan!10}88.2\% \\ 
\midrule
\multirow{3}{*}{\shortstack{IRAF\\(\textit{Proposed})}} 
 & LIBRI & 11.64 & 0.472 & 0.94 & 91.2\% \\
 & US-GOV & 12.34 & 0.486 & 0.93 & 92.8\% \\
 & \cellcolor{cyan!10}ALL 
 & \cellcolor{cyan!10}\textbf{12.01} 
 & \cellcolor{cyan!10}\textbf{0.476} 
 & \cellcolor{cyan!10}\textbf{0.94} 
 & \cellcolor{cyan!10}\textbf{92.5\%} \\ 
\bottomrule
\end{tabular}}
\vspace{-0.2cm}
\end{table}

\begin{table*}[htbp]
\centering
\caption{Performance on InstructS2S-200K under MUSAN speech interference. $\Delta$ (IRAF-NoisyAug) with blue boxes shows gains of the proposed IRAF over NoisyAug; values in parentheses denote relative improvements.}
\vspace{-0.3cm}
\label{tab2:instruct200k}
\resizebox{0.9\textwidth}{!}{
\begin{tabular}{lcccccc}
\toprule
\multirow{2}{*}{\textbf{Method}} &
\multicolumn{2}{c}{\textbf{Response Quality}} &
\multicolumn{2}{c}{\textbf{Turn-taking Performance}} &
\multicolumn{2}{c}{\textbf{Barge-in Performance}} \\ 
\cmidrule(lr){2-3} \cmidrule(lr){4-5} \cmidrule(lr){6-7}
& \textbf{BLEU($\uparrow$)} & \textbf{sBERT($\uparrow$)}
& \textbf{RL(s)($\downarrow$)} & \textbf{RSR($\uparrow$)} 
& \textbf{SL(s)($\downarrow$)} & \textbf{SSR($\uparrow$)} \\ 
\midrule
\multicolumn{6}{l}{\textit{Interfering speakers only}} \\ 
\midrule
CleanBase        & 1.13 & 0.22 & 1.39 & 13.9\% & 1.29 & 42.7\% \\
NoisyAug         & 9.64 & 0.47 & 0.97 & 69.2\% & 0.74 & 99.0\% \\
IRAF (\textit{Proposed})  & \textbf{13.76} & \textbf{0.58} & \textbf{0.82} & \textbf{91.0\%} & \textbf{0.73} & \textbf{99.8\%} \\ 
\rowcolor{cyan!10} $\Delta$ (IRAF-NoisyAug)  & +4.12 {\scriptsize (+42.73\%)} & +0.11 {\scriptsize (+23.40\%)} & -0.15 & +21.8\% & -0.01 & +0.8\% \\ 
\midrule
\multicolumn{6}{l}{\textit{Both interfering speakers and background noise}} \\ 
\midrule
CleanBase        & 0.91 & 0.21 & 1.41 & 9.8\% & 1.34 & 40.2\% \\
NoisyAug         & 8.32 & 0.44 & 1.05 & 56.0\% & 0.74 & 99.6\% \\
IRAF (\textit{Proposed})  & \textbf{9.83} & \textbf{0.47} & \textbf{0.98} & \textbf{69.2\%} & \textbf{0.73} & \textbf{100.0\%} \\ 
\rowcolor{cyan!10} $\Delta$ (IRAF-NoisyAug)   & +1.51 {\scriptsize (+18.15\%)} & +0.03 {\scriptsize (+6.81\%)} & -0.07 & +13.2\% & -0.01 & +0.4\% \\
\bottomrule
\end{tabular}}
\vspace{-0.5cm}
\end{table*}

\vspace{-0.3cm}
\section{Experiments}

\subsection{Experimental Setup}
\vspace{-0.1cm}
The model is implemented using the NeMo Toolkit~\cite{kuchaiev2019nemotoolkitbuildingai}. The speech encoder is initialized from a 100M-parameter streaming pretrained encoder with an 80\,ms right context~\cite{speech_encoder}, and the LLM is initialized from the 1.1B-parameter TinyLlama model~\cite{zhang2024tinyllamaopensourcesmalllanguage}. For speech, we adopt NanoCodec~\cite{casanova2025nanocodechighqualityultrafast} at 0.6\,kbps by default. The resulting speech representation comprises four code channels, each with a vocabulary size of 4,037. The speech decoder is a 12-layer causal Transformer following T5 architecture~\cite{raffel2020exploring}. For text, we use a 32k SentencePiece tokenizer. In the IRAF module, the Causal Transformer has only 1 transformer layer. We extract target-speaker embeddings using a pretrained ECAPA-TDNN~\cite{desplanques20_interspeech} and, for simplicity, treat speakers from different conversations as distinct identities. During training, AdamW with a cosine-annealing learning rate scheduler is performed. The peak learning rate is \(3\times 10^{-4}\), with 2,500 warm-up steps. We apply gradient clipping with a maximum norm of 1.0 to stabilize training. Each dataset is partitioned into train, validate, and test splits with ratios of 0.945, 0.005, and 0.05, respectively. For MS MARCO, we use a per-GPU batch size of 1 and set accumulate grad batches as 8. For InstructS2S-200K, we employ duration-based bucketing with a batch duration of 60\,s and set accumulate grad batches to 4 per GPU.

\vspace{-0.15cm}
\subsection{Evaluation Metrics}
\vspace{-0.1cm}
We evaluate the proposed model from two complementary perspectives: a) response quality to assess reasoning and answer correctness, and b) full-duplex interaction to assess turn-taking and barge-in behavior.

\noindent\textbf{Response quality}: We transcribe the generated agent speech using an ASR system~\cite{asr_score}, and then compute the BLEU score~\cite{blue_metrics} and the Sentence-BERT (sBERT) semantic similarity~\cite{reimers-gurevych-2019-sentence} between the ASR transcript and the reference text.

\noindent\textbf{Turn-taking performance}: We report response latency (RL) and response success rate (RSR). Using Silero VAD~\cite{Silero_VAD}, we estimate the user end-of-speech and the agent speech onset; response latency is the time between them. Latencies exceeding 1.5 s are treated as failures and clipped to 1.5 s when reporting.


\noindent\textbf{User barge-in performance}: For multi-turn conversations, we use Barge-in stop latency (SL) and stop success rate (SSR). We define stop latency as the time between the user’s interruption onset and the time when the agent ceases speaking (estimated via VAD). As above, cases with latency larger than 1.5s are treated as failures and are clipped to 1.5s for latency reporting. To increase the likelihood of interruptions at test time, we construct a barge-in evaluation set by shortening the gaps between consecutive user turns, thereby reducing the time available for the agent to complete its response.

\vspace{-0.3cm}
\subsection{Results Analysis}
\vspace{-0.15cm}
We report results under three configurations. CleanBase serves as the primary baseline, while NoisyAug reflects a commonly used robustness strategy; \textbf{a) CleanBase}: E2E full-duplex model without IRAF, trained on clean speech. \textbf{b) NoisyAug}: the same model without IRAF, trained with noise/interference augmentation. \textbf{c) IRAF}: the proposed E2E full duplex model with IRAF.

Table~\ref{tab:msmarco_people} highlights three consistent trends on the single-turn MS-MARCO conversation dataset. \textbf{a)} CleanBase, trained only on clean speech, shows clear degradation in reasoning-oriented response quality (BLEU and sBERT) under noisy/interfered test conditions. The low response success rate further indicates that the model often fails to respond promptly after the user finishes speaking, suggesting that noise/interference disrupts the model’s turn-taking and response triggering. \textbf{b)} When we inject interfering speakers during training on-the-fly (Sec. 3), the model improves both response quality (e.g., BLEU reaching 12.74) and interaction reliability (success rate 93.1\%). These gains are largely attributable to exposure to overlap and non-target speech during training, which reduces distribution mismatch and improves the model’s ability to detect user end-of-speech under interference. \textbf{c)} The proposed IRAF module yields further gains beyond augmentation, improving BLEU by +1.46 (+11.46\% relative) and success rate by +2.6\% absolute, consistent with IRAF’s goal of suppressing interference-induced conditioning corruption via frame-level reliability gating. The same pattern holds in the more extreme setting that combines interfering speakers with background noise, where IRAF maintains similarly significant improvements.

\begin{figure}[!ht]
  \centering
  \includegraphics[width=0.8\linewidth]{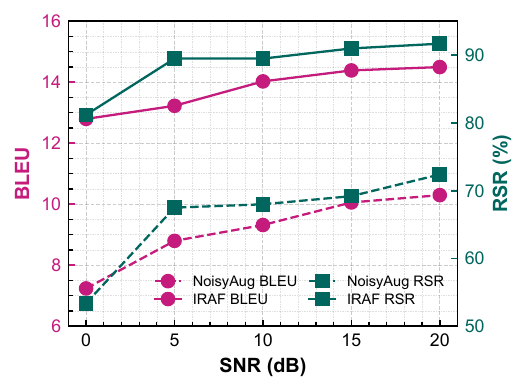}
  \vspace{-0.5cm}
  \caption{BLEU and response success rate (RSR) on InstructS2S-200K with interfering speakers across SNRs.}
  \label{fig:snr}
  \vspace{-0.2cm}
  
\end{figure}

We further evaluate IRAF on a multi-turn InstructS2S-200K conversational benchmark to reflect more realistic, long-horizon interactions. The main trends from the single-turn setting persist, but the multi-turn scenario amplifies the interaction-control difficulty, as shown in Table~\ref{tab2:instruct200k}. \textbf{a)} The proposed IRAF consistently improves response quality, achieving +4.12 BLEU (+42.7\% relative) and +0.11 sBERT (+23.4\% relative), indicating better semantic faithfulness under sustained noise. \textbf{b)} In multi-turn dialogue, naive interference augmentation yields a low response success rate (69.2\%), indicating that simply mixing interfering speakers during training does not adequately address missed responses. With IRAF, response success increases to 91\% and response latency drops to 820 ms, demonstrating improved reliability and faster responses under interference. \textbf{c)} IRAF also supports accurate interruption handling: it achieves a 730 ms stop latency and 99.8\% stop success rate when the user barges in under the interfering-speaker-only scenario, and these strong results hold even for the more extreme interfering-speaker with background-noise conditions. \textbf{d)} In addition, as shown in Figure~\ref{fig:snr}, IRAF consistently improves both audio reasoning quality and response success rate across all SNR settings, with gains that remain stable under different interfering-speaker conditions. These results indicate that IRAF generalizes across a broad range of interference severities, rather than being tuned to a particular SNR regime. Overall, IRAF enables responses that are faster, more accurate in turn-taking, and higher in quality in multi-turn settings.

\vspace{-0.4cm}

\vspace{-0.5cm}
\section{Conclusions} \label{sec:conclusion}
\vspace{-0.2cm}
To address the key challenge of interference-induced conditioning corruption in end-to-end full-duplex spoken dialogue systems, this paper presented IRAF, a lightweight, streaming-compatible adaptive fusion module that performs frame-level reliability gating using target-speaker and user-audio embeddings before fusion with agent representations. Experiments on MS-MARCO and InstructS2S-200K confirmed consistent improvements in both response quality and full-duplex interaction under interfering-speaker conditions, demonstrating IRAF’s effectiveness across a range of interference levels.

\bibliographystyle{IEEEtran}
\bibliography{new_mybib}

\end{document}